\documentclass[mathleft]{an}

\usepackage{graphicx}
\usepackage{times}
\usepackage{amsmath}
\usepackage{mathabx}
\usepackage{ctable}

\begin{document}

\Pagespan{1}{}

\title{X-shooter observations of QSO pairs\thanks{Based on observations collected at the European Southern Observatory Very Large Telescope, Cerro Paranal, Chile -- Instrument commissioning and program 084.A-0252(A).}}

\author{Guido Cupani\inst{1}\fnmsep\thanks{Corresponing author: \email{cupani@oats.inaf.it}\newline}
\and Valentina D'Odorico\inst{1}
\and Stefano Cristiani\inst{1,2}
\and Matteo Viel\inst{1}
\and Eros Vanzella\inst{1}
}
\titlerunning{X-shooter observations of QSO pairs}
\authorrunning{G.~Cupani\inst{1}}
\institute{INAF-OATs, via Tiepolo 11, 34131 Trieste, Italy
\and
INFN/National Institute of Nuclear Physics, via Valerio 2, 34127 Trieste, Italy
}


\abstract{Observations of quasar pairs represent a powerful tool in the study of the intergalactic medium, providing in particular information on the quasar radiative feedback not only along the line of sight, but also in the transverse direction. 
In this paper, we present the spectra of the four quasar pairs we observed so far with X-shooter, and we discuss two individual spectra affected by the presence of a close quasar in the foreground.}

\maketitle

\section{Introduction: QSO pairs}

In the last decades, the analysis of Lyman-$\alpha$ and metal-line absorbers in the spectra quasars (QSOs) has dramatically improved our knowledge of the intergalactic medium (IGM) at intermediate redshifts ($z\sim 2$--$4$). 
Regrettably, observations of isolated quasars provides information only along the line of sight.
To overcome this limitation, a suitable choice is to observe pairs and groups of quasar at small angular separation ($\sim$$1$--$10^\prime$), which allow us to obtain information in the transverse direction, and 
assess the redshift space distortion due to IGM kinematics.

At present, suitable samples of spectra of QSO pairs are required in many fields. 
Coincidences of absorption features along paired sightlines are useful to assess the size of both Ly-$\alpha$ absorbers (e.g., D'O\-dorico et al. 1998) and metal-line absorbers (e.g., Martin et al. 2010), constraining the IGM geometry and its physical and chemical state. 
Pairs with large redshift separations make it possible to investigate the radiative feedback from quasars in 3D, revealing in the spectrum of the background quasar the local enhancement of the photoionization rate due to the foreground quasar (the so-called transverse proximity effect, or TPE; e.g., Crotts 1989; Fern\'andez-Soto et al. 1995; Gon\c calves, Steidel \& Pettini 2008).
In addition, accurate measurements of the auto- and cross-correlation of the Ly-$\alpha$ forest are expected to provide an independent estimation of the cosmological parameter $\Omega_\Lambda$, via an adapted version of the Alcock-Paczy\'nski (AP) test (e.g., McDonald 2003). 
Such analysis requires simulations to properly reconstruct the 3D density field as a function of cosmology (e.g., Pichon et al. 2001).
The samples of QSO pairs collected so far (e.g., Coppolani et al. 2006, D'Odorico et al. 2006) are too small to perform 
the AP 
test.

In this framework the spectrograph X-shooter (see contributions by S. D'Odorico, J. Vernet, and F. Zerbi) represents a breakthrough. 
Its intermediate resolution, extended spectral coverage, and high efficiency make it an ideal instrument to collect spectra of QSO pairs that would be too faint for the high-resolution spectrographs at the $8$--$10$ m class telescopes. 
In this paper, we describe the present state of the X-shooter GTO program dedicated to the observation of QSO pairs and discuss two individual (and significantly different) cases of TPE in the spectra extracted so far. 

\section{The X-shooter sample}

Four QSO pairs have been observed so far with X-shooter. 
The first one (2QZ J0202--2805 and J0202--2804, at an angular separation $\Delta\theta=47.2^\second$) was already targeted in No\-vember, 2008, during the instrument commissioning. 
We observed three more pairs (CTS 426 and 427, $\Delta\theta=58.2^\second$; SDSS J0111+1401 and J0111+1402, $\Delta\theta=35.8^\second$; 2QZ J0306--3010 and J0306\- --3011, $\Delta\theta=51.5^\second$) in October, 2009, within the GTO program. All quasar pairs have emission redshift $z\sim 2$--$3$ (Maza et al.~1995; Croom et al.~2001; Croom et al.~2004; Marble et al.~2008). 
These observations were reduced into 2D spectra with the public release X-shooter pipeline (see contributions by A. Modigliani and P. Goldoni); we used the ESO-MIDAS package to extract the 1D spectra from the 2D ones. 
The typical signal-to-noise ratio per wavelength bin (0.3 \AA) of the resulting spectra is $\sim$$30$--$60$ at the peak of the Ly-$\alpha$ emission, and $\sim$$10$--$40$ in the Ly-$\alpha$ forest and in the C \textsc{iv} forest. 
These spectra provide a significative addition to the sample of quasar pairs and groups already observed with UVES (D'Odorico et al. 2006, Cappetta et al. 2010). 
In the context of the GTO program, we expect to obtain $\sim$11 pairs, which combined with the pairs observed with UVES add up to a total sample of $\sim$27 
pairs. 
We are going to perform new simulations in order to assess the significance of the result we will be able to obtain with this sample, which according to McDonald (2003) should be 
large enough to perform the AP test 

\begin{figure}
\center
\resizebox{0.90\hsize}{!}{\includegraphics[angle=270]{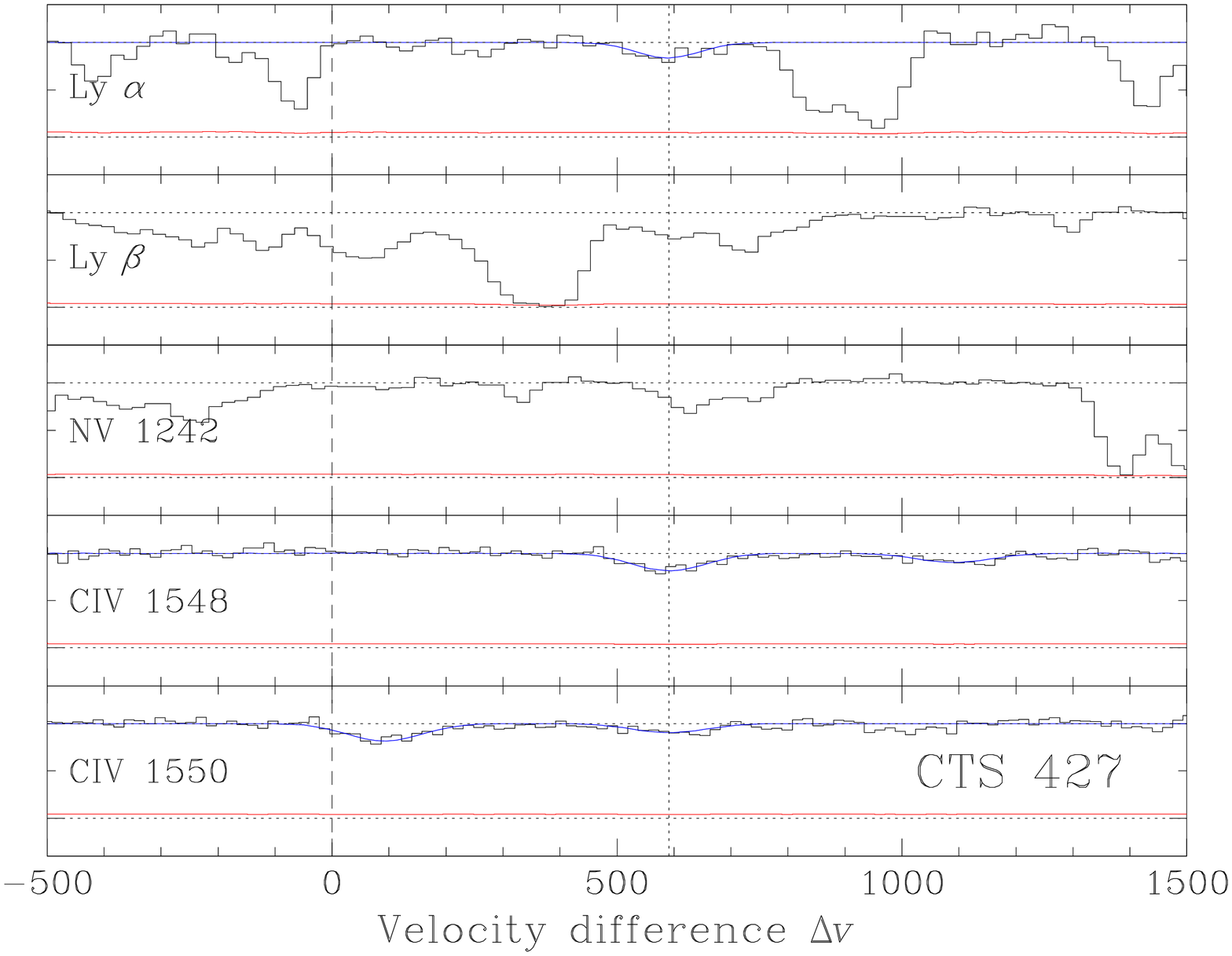}}
\resizebox{0.90\hsize}{!}{\includegraphics[angle=270]{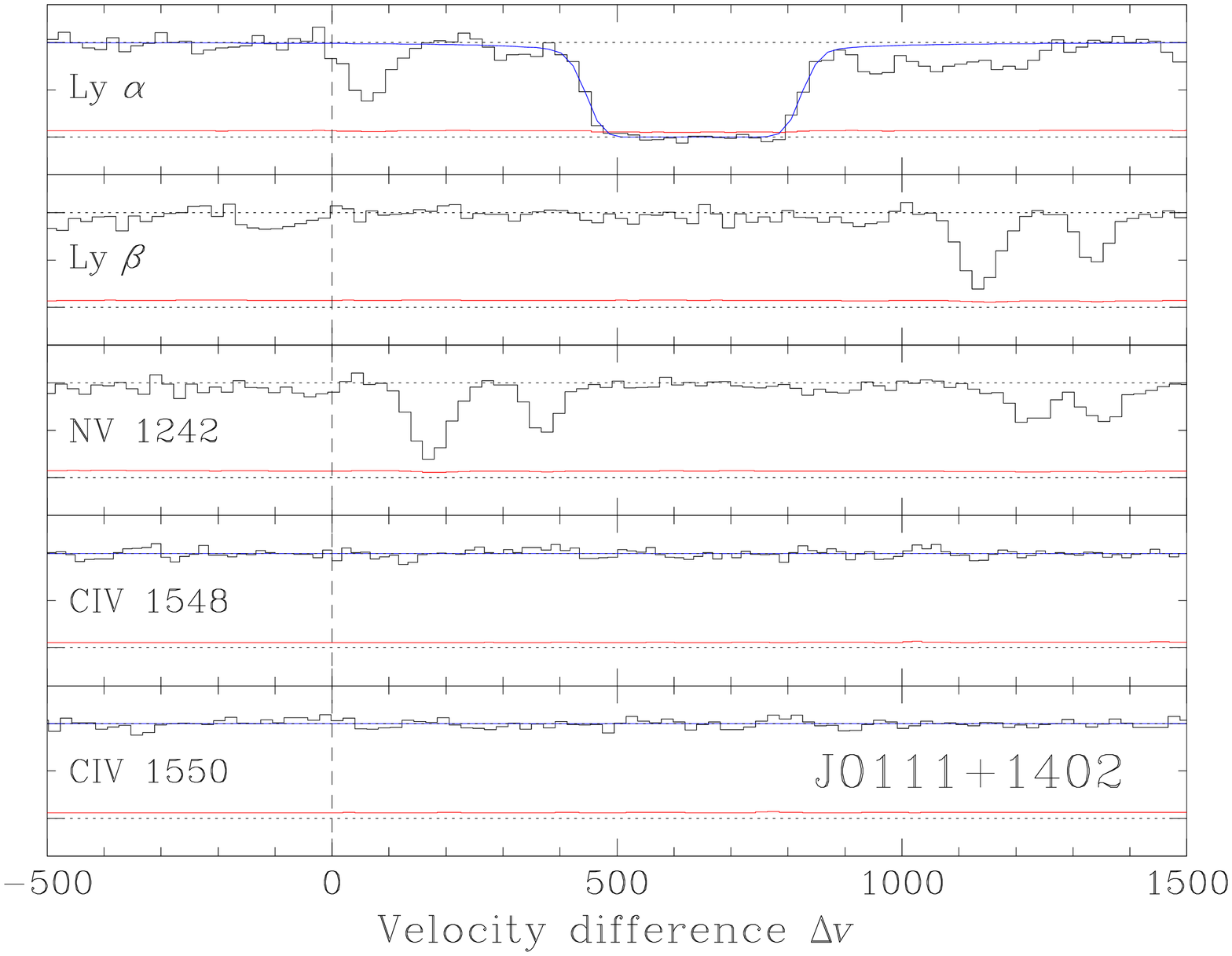}}
\caption{Regions of the background QSOs spectra around the foreground QSOs redshift. Upper panel: spectrum of CTS 427; lower panel: spectrum of J0111+1402. Normalized flux and error are shown in black and red, respectively. The blue line is a parametric fit of the absorption lines obtained with \texttt{FIT/LYMAN} in ESO-MIDAS.}
\label{fig:specfit}
\end{figure}

Only in two cases (pairs 2 and 3) the redshift difference between the foreground and the background QSO is large enough to allow a positive assessment of the TPE. Fig.~\ref{fig:specfit} shows the regions of the background QSOs were the absorption signatures of the foreground QSOs (Ly-$\alpha$ line, N \textsc{v} $\lambda\lambda 1238, 1242$ and C \textsc{iv} $\lambda\lambda 1548, 1550$) are expected to be found. The following paragraphs describes the two pairs in further details.
We remark that TPE is by nature a statistical effect, so conclusions based on individual pairs should be regarded as tentative.

\paragraph{CTS 426 and 427.} The velocity difference between the two QSOs along the line of sight is $\Delta v\sim 10500$ km s$^{-1}$, corresponding to a proper distance of $\sim$$50h^{-1}$ Mpc. 
The regions of interest of the spectrum of CTS 427 ($z_\textrm{em}=2.33\pm 0.02$) are shown in the upper panel of fig.~\ref{fig:specfit}. 
Although no evident absorption is found at the redshift of CTS 426 ($z_\textrm{em}=2.215\pm 0.0044$, corresponding to $\Delta v=0$), we detected a system just $\sim$$590$ km s$^{-1}$ redwards (dotted vertical line), with very weak Ly-$\alpha$ absorption ($\log N_\textrm H=13.26\pm 0.08$) and evidence of consistent ionization ($\log N_\textrm{C \textsc{iv}}=13.62\pm 0.04$; possible N \textsc{v} doublet blended with other lines). 
We interpret this observation as a signature of TPE in this pair.

\paragraph{SDSS J0111+1401 and J0111+1402.} This case is quite the opposite of the previous one. 
The redshift difference between the two QSOs is the highest in our sample ($\Delta z\sim 0.45$).
Several Ly-$\alpha$ lines are observed in the spectrum of J0111+1402 ($z_\textrm{em}=2.93448\pm 0.001778$) around the redshift of J0111+1401, and a remarkable saturated absorption feature is recognizable approximately $650$ km s$^{-1}$ redwards the systemic redshift of J0111+1401, $z_\textrm{em}=2.48\pm 0.17$, estimated by us  from the O \textsc{i} $\lambda 1302$ line (a slightly lower value, $z_\textrm{em}=2.47$, has been published by Hennawi et al. 2006a). 
The lower panel of fig.~\ref{fig:specfit} uses our new redshift value to define $\Delta v$. 
The saturated Ly-$\alpha$ absorption is compatible with the presence of J0111+1401 at a proper distance, neglecting peculiar motions, of $\sim$$3.5h^{-1}$ Mpc (the transverse separation of J0111+1401 with respect to the line of sight to J0111+1402 is $\sim$$290h^{-1}$ kpc), and is associated to a lack of detected metal lines. 
In this case, the enhanced ionization expected as a consequence of TPE is like\-ly masked by an anisotropic distribution of gas, due to clustering around the foreground QSO (see, e.g., Fern\'andez-Soto et al. 1995; Hennawi et al. 2006b).


\end{document}